\documentstyle[prl,twocolumn,aps,tighten,floats,epsfig]{revtex}
\begin{document}

\narrowtext

{\bf Comment on ``Origin of Giant Optical Nonlinearity in 
Charge-Transfer--Mott Insulators: A New Paradigm for Nonlinear Optics''}
\medskip

Zhang \cite{Zhang} has proposed a novel mechanism for the 
giant optical nonlinearities in linear chain Nickel Halides (Ni-X)
\cite{Kishida} and the cuprate 
Sr$_2$CuO$_3$ \cite{Ogasawara}. We show that Zhang's theory is inapplicable to 
these systems and also, some of his numerical results are finite size artifacts.
Although Zhang's Hamiltonian for Ni-X is correct, the nature of the
eigenstates depend crucially on the {\it relative} site energies of Ni and X 
in $\Delta \sum_{i,\sigma}(-1)^in_{i,\sigma}$. 
Zhang chose the Ni atoms to occupy 
the odd sites in his exact diagonalization of N = 8
sites, such that the site energy of Ni is {\it lower} than that of X. There
is then a competition between the on-site correlation $U_{Ni}$ 
(which prefers Ni$^{3+}$) and
$\Delta$ (which prefers Ni$^{2+}$), and near the interface, 
the nonlinear coefficient $\gamma$ is huge for $\Delta < \Delta_c(U)$
and zero for $\Delta > \Delta_c(U)$ (see Fig.~1 of \cite{Zhang}). 
Based on earlier estimates of 
$U$ and the {\it magnitude} of $\Delta$, Zhang claims that Ni-X lie very
close to the interface.

Zhang's choice of the relative site energies
is {\it opposite} to that of {\it all} previous authors \cite{Roder,Okamoto},
and can be precluded.
Within the Ni$^{3+}$ phase, the charge-transfer (CT) gap with conventional 
parametrization \cite{Roder,Okamoto} increases with $(U_{Ni} - U_X) + V + 
2\Delta$, whereas with Zhang's choice the gap increases with
$(U_{Ni} - U_X) + V - 2\Delta$.
With a less diffuse valence atomic orbital in Cl, $U_{Cl} \geq U_{Br}$.
Since, however, Cl is more electronegative than Br, 2$\Delta$ is larger for 
X = Cl with conventional parametrization \cite{Roder,Okamoto} and
compensates for the smaller
$(U_{Ni} - U_X)$ to give a larger optical gap in Ni-Cl, as is experimentally
observed \cite{Okamoto}. With Zhang's choice of site energies, 
the only way to explain the larger CT gap in Ni-Cl is to make the unphysical 
assumption that 2$\Delta$ is smaller for Cl than Br, i.e.,
Cl is less electronegative than Br!
Thus the conventional parametrization is correct.
Hence there is no competition between
$U_{Ni}$ and $\Delta$, and no enhancement of $\gamma$.

A second error in \cite{Zhang} involves the the sudden drop in $\gamma$
by orders of magnitude to zero (see Fig.~1 of \cite{Zhang}). We show that this 
is a consequence of
a crossover of the ground state from total spin S = 0 to S = 1, a finite size
effect. It is 
well known that the ground state of a {\it finite} undistorted non-half-filled
periodic ring with 4n electrons  
\begin{figure}
\centerline{\epsfig{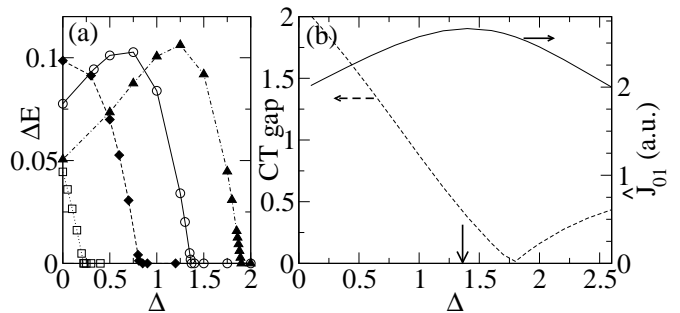}}
\smallskip
\caption{(a) Energy difference between the lowest S$_z$ = 1 and  
S$_z$ = 0 states of the N = 8 periodic ring with 12 electrons, for, 
from left to right,
$U_{Ni}$ = 2, 3, 4 and 5, with other parameters the same as in 
Fig.~1 of \protect\cite{Zhang}. 
(b) The CT gap and the matrix element of the current operator $\hat{J}$
for S = 0 and $U_{Ni}$ = 4. The arrow indicates the $\Delta$ where 
spin crossover occurs.}
\end{figure}
is S = 1. With $U_{Ni} \neq U_X$ and $V > 0$ this occurs
at nonzero $\Delta$. Zhang must have used a numerical 
approach that conserves total S$_z$ and not total S. In Fig.~1(a) we 
have plotted
$\Delta$E = E(S$_z$ = 1) -- E(S$_z$ = 0) for the parameters of Zhang's
Fig.~1. In all cases $\Delta$E vanishes at {\it exactly} the same $\Delta_c$ 
as in \cite{Zhang}. Using a method that conserves
total S, we have confirmed that for $\Delta < \Delta_c$, E(S = 0) = E(S$_z$ = 0),
while for $\Delta > \Delta_c$, E(S = 0) $>$  E(S$_z$ = 0). With specifically
N = 8 and 12 electrons, the S = 1 ground state is not coupled
to excited states by the current operator, and this 
is the reason for the sharp drop in $\gamma$ in \cite{Zhang}. We have confirmed 
that the spin crossover does not occur for N = 12 with
18 electrons. Also,
the true $\Delta_c$ that defines the Ni$^{3+}$--Ni$^{2+}$ interface is larger for
all $U$. 
In Fig.~1(b) we have plotted the CT gap as well as the matrix element of the
current operator $\hat{J}_{01}$ between the lowest S = 0 state and 
the S = 0 one-photon
state for the N = 8 periodic ring, using the site energies in \cite{Zhang}.
$\hat{J}_{01}$ is symmetric about the true $\Delta_c$
indicating a nearly symmetric behavior of $\gamma$ 
even with Zhang's parametrization.

To conclude: (i) Zhang's choice of site energies for Ni-X are incorrect, (ii)
his determination of $\Delta_c$ is incorrect, and (iii) the calculated behavior
of $\gamma$ for $\Delta > \Delta_c$ in \cite{Zhang}
is an artifact of finite size calculations.
Finally, the giant increase in $\gamma$ 
near $\Delta_c(U)$ \cite{Zhang} within the model 
is largely due to decreasing CT gap rather than
increasing $\hat{J}_{01}$.
This does not permit device application, as losses due
to absorption at wavelengths of interest 
would be large. We acknowledge useful discussions with A. Painelli and partial 
support from NSF-DMR.
\medskip

R.T. Clay$^{1,2}$ and S.~Mazumdar$^1$

$^1$ Department of Physics, University of Arizona, 
\indent Tucson, AZ 85721

$^2$ Cooperative Excitation Project ERATO, Japan 
\indent Science and Technology Corporation (JST), University
\indent of Arizona, Tucson, AZ 85721

\end{document}